\documentclass[a4paper,11pt]{article}
\pdfoutput=1
\usepackage{jcappub}
\usepackage[T1]{fontenc}

\usepackage{multirow}
\usepackage{amssymb}
\usepackage{floatrow}
\usepackage{amsmath}
\usepackage{aas_macros}
\usepackage{graphicx}
\usepackage{color}
\usepackage{array}
\usepackage{varwidth}
\usepackage{hyperref}
\hypersetup{
    pdfnewwindow=true,
    colorlinks=true,
    linkcolor=blue,
    citecolor=blue,
    filecolor=blue,
    urlcolor=blue
}

\title{The Highest Energy HAWC Sources are Likely Leptonic and Powered by Pulsars}

\author[a,1]{Takahiro Sudoh,\note{Corresponding author.}}
\author[b]{Tim Linden,}
\author[c,d]{Dan Hooper}

\affiliation[a]{Department of Astronomy, University of Tokyo, Hongo, Tokyo 113-0033, Japan}
\affiliation[b]{Stockholm University and The Oskar Klein Centre for Cosmoparticle Physics, Alba Nova, 10691 Stockholm, Sweden}
\affiliation[c]{Theoretical Astrophysics Group, Fermi National Accelerator Laboratory, Batavia, Illinois, 60510, USA}
\affiliation[d]{Department of Astronomy and Astrophysics and the Kavli Institute for
Cosmological Physics (KICP), University of Chicago, Chicago, Illinois, 60637, USA}

\emailAdd{sudoh@astron.s.u-tokyo.ac.jp}
\emailAdd{linden@fysik.su.se}
\emailAdd{dhooper@fnal.gov}

\def\lsim{\mathrel{\raise.3ex\hbox{$<$\kern-.75em\lower1ex\hbox{$\sim$}}}}
\def\gsim{\mathrel{\raise.3ex\hbox{$>$\kern-.75em\lower1ex\hbox{$\sim$}}}}

\abstract{The HAWC Collaboration has observed gamma rays at energies above 56 TeV from a collection of nine sources. It has been suggested that this emission could be hadronic in nature, requiring that these systems accelerate cosmic-ray protons or nuclei up to PeV-scale energies. In this paper, we instead show that the spectra of these objects favor a leptonic (inverse Compton) origin for their emission. More specifically, the gamma-ray emission from these objects can be straightforwardly accommodated within a model in which $\sim \mathcal{O}(10\%)$ of the host pulsar's spindown power is transferred into the acceleration of electrons and positrons with a power-law spectrum that extends to several hundred TeV or higher. The spectral break that is observed among these sources is naturally explained within the context of this simple model, and occurs at the energy where the timescale for energy losses matches the age of the pulsar. In contrast, this spectral feature cannot be straightforwardly accommodated in hadronic scenarios. Furthermore, hadronic models predict that these sources should produce more emission at GeV-scale energies than is observed. In light of these considerations, we conclude that HAWC's highest energy sources should be interpreted as TeV halos or pulsar wind nebulae, which produce their emission through inverse Compton scattering, and are powered by the rotational kinetic energy of their host pulsar.}

\begin{document}
\maketitle

\section{Introduction}

The cosmic-ray spectrum is thought to be dominated by Galactic sources up to energies of $\sim$\,1 PeV, corresponding to the spectral feature known as the ``knee''. The nature of the Milky Way's so-called ``PeVatrons'' remains an open and widely debated question. Among the proposed candidates, supernova remnants have long been the most popular, and gamma-ray measurements support the conclusion that these objects produce high-energy protons~\cite{2010ApJ...710L.151T,Ackermann:2013wqa}. That being said, it has also been argued that such sources may be unable to accelerate protons beyond a few hundred TeV~\cite{Bell:2013kq,Gabici:2016fup}. Other PeVatron candidates include the Milky Way's supermassive black hole~\cite{Fujita:2016yvk,Guo:2016zjl,Abramowski:2016mir}, and clusters of young and massive stars~\cite{Aharonian:2018oau}. The sources of the highest energy Galactic protons are expected to generate gamma rays through the production and decay of neutral pions, resulting in a power-law gamma-ray spectrum that extends to $\sim$100 TeV.

Pulsars can also accelerate electrons and positrons up to energies of at least $\sim$100 TeV. Due to the Klein-Nishina suppression associated with inverse-Compton scattering, electrons and positrons in this energy range lose much of their energy to synchrotron emission, suppressing the leptonic production of $\sim$100 TeV-scale gamma rays. Through this distinction, very high-energy gamma-ray telescopes provide us with one of the most powerful ways to discriminate between accelerators of hadronic and leptonic cosmic rays.

The High Altitude Water Cherenkov (HAWC) observatory has recently produced a catalog of nine gamma-ray sources detected at energies above 56 TeV. Three of these sources have been observed above 100 TeV, making this the highest energy gamma-ray catalog reported to date~\cite{Abeysekara:2019gov}.\footnote{The Tibet air shower array has also reported the detection of emission above 100 TeV from the Crab Nebula~\citep{2019PhRvL.123e1101A}.} Given that all nine of these sources are located within $0.5^{\circ}$ of a known pulsar, it appears likely that they are associated with this class of objects. Furthermore, eight of these nine pulsars are quite young ($t_{c}\equiv P/2\dot{P} \sim 1-50 \, {\rm kyr}$), and have exceptionally high spindown power ($\dot{E} > 10^{36} \, {\rm erg/s}$). This information suggests two possible interpretations. On the one hand, the gamma-ray emission from these sources could be leptonic in nature, powered by the host pulsars' rotational kinetic energy. Alternatively, the observed emission could be hadronic, revealing these systems' supernova remnants to be among the Milky Way's long-sought-after PeVatrons.

In this paper, we examine the luminosity, spectrum, and morphology of the very high-energy sources observed by HAWC in order to evaluate whether they are more likely to be leptonic sources powered by the rotational kinetic energy of the young pulsar, or hadronic PeVatrons powered by the supernova remnant. We find that the former interpretation is favored by three factors. First, the spectra of these sources can be easily accommodated by simple models in which very high-energy electrons and positrons are accelerated with a power-law spectrum. In contrast, hadronic models cannot straightforwardly account for the spectra observed from several of HAWC's highest energy sources. Second, the gamma-ray luminosities observed from these sources are well-matched to the past integrated spindown power of their host pulsars. And third, the spectral break observed among these systems at $E_{\gamma} \sim \mathcal{O}(10 \,{\rm TeV})$ is naturally explained by the guaranteed suppression of the inverse Compton scattering cross section by Klein-Nishina effects, and the energy dependence of the electron/positron energy-loss time-scale, which is smaller than the pulsar age for the highest-energy leptons.

In light of these considerations, we conclude that HAWC's highest energy sources are likely to be TeV halos and/or pulsar wind nebulae, with gamma-ray emission that is 1) powered by the rotational kinetic energy of the host pulsar, and 2) produced through inverse Compton scattering.

\section{TeV Halos and Pulsar Wind Nebulae}

Observations by HAWC and Milagro have detected diffuse multi-TeV emission from the regions surrounding the nearby Geminga and Monogem pulsars~\cite{Albert:2020fua,Abeysekara:2017hyn,Abeysekara:2017old,Abdo:2009ku}. The spectrum and intensity of this emission indicate that these sources convert a significant fraction ($\sim$\,$10\%$) of their total spindown power into very high-energy electron-positron pairs. Furthermore, each of these TeV halos exhibits an angular extension of $\sim$\,$2^{\circ}$ (corresponding to $\sim$\,$25 \,{\rm pc}$), indicating that cosmic-ray propagation in the vicinity of these pulsars is much less efficient than is typically experienced elsewhere in the interstellar medium~\cite{Hooper:2017gtd,Hooper:2017tkg,2018MNRAS.479.4526L,Johannesson:2019jlk,DiMauro:2019hwn,Liu:2019zyj,Evoli:2018aza, Kun:2019sks}.

Looking beyond the specific examples of Geminga and Monogem, observations by HAWC (and HESS~\cite{H.E.S.S.:2018zkf,Abdalla:2017vci}) have led to the identification of a new class of spatially extended, multi-TeV gamma-ray sources, powered by the rotational kinetic energy of pulsars, and which produce their observed emission through the inverse Compton scattering of very high-energy electrons and positrons on the surrounding radiation field~\cite{Linden:2017vvb,Sudoh:2019lav}. A large fraction of the sources detected by HAWC~\cite{Albert:2020fua,Abeysekara:2017hyn,Smith:2020clm} have been shown to be spatially coincident with a pulsar, and all indications suggest that TeV halos are a generic feature of middle-aged pulsars (whether or not TeV halos also accompany millisecond pulsars is an open question~\cite{Hooper:2018fih}). These observations suggest that nearby TeV-halos are likely responsible for the observed cosmic-ray positron excess~\cite{Hooper:2017gtd,Fang:2018qco,Profumo:2018fmz,Fang:2018qco,Tang:2018wyr,Manconi:2020ipm} (for earlier work, see Refs.~\cite{2009PhRvL.103e1101Y,1995A&A...294L..41A,1995NuPhS..39..193A}), as well as the diffuse TeV excess observed by Milagro~\cite{Linden:2017blp}, and could plausibly dominate the TeV-scale emission observed from the Galactic Center by HESS~\cite{Hooper:2017rzt} (as opposed to the hypothesis that this emission is produced by a Galactic Center PeVatron~\cite{Abramowski:2016mir}). Extrapolating to the Milky Way's larger pulsar population, we expect HAWC and the Cherenkov Telescope Array (CTA)~\cite{CTAConsortium:2018tzg} to ultimately detect $\sim$\,$50-240$ TeV halos~\cite{Sudoh:2019lav}, including many whose pulsed radio and gamma-ray emission is not beamed in the direction of Earth~\cite{Linden:2017vvb}.

When referring to TeV halos, we adopt a definition for this source class which requires that the high-energy electrons and positrons responsible for the observed gamma-ray emission propagate via diffusion, rather than convection or advection. This distinguishes TeV halos from pulsar wind nebulae, for which advection plays an important and often dominant role (for a review, see Ref.~\cite{Gaensler:2006ua}). TeV halos are also more spatially extended than typical pulsar wind nebulae. Pulsar wind nebulae are created when the energetic outflow from a pulsar collides with the ambient medium (supernova ejecta or interstellar medium), resulting in a shockwave surrounding a diffuse plasma of electrons and positrons. Like TeV halos, the emission from a pulsar wind nebula is powered by its pulsar's rotational kinetic energy, and is leptonic in nature. We consider it to be plausible that HAWC's highest energy sources could be a combination of TeV halos, pulsar wind nebulae, and objects that are currently in a transitional state between these two classifications.\footnote{An alternative definition has been put forth by Giacinti {\it et al.}~\cite{Giacinti:2019nbu} which classifies a region as a TeV halo if it contains an overdensity of relativistic electrons and positrons around a pulsar, and if the pulsar and associated supernova remnant does not dominate the dynamics or composition of the interstellar medium in that region. Compared to our definition, this choice leads Giacinti~{\it et al.} to classify many objects that we would call TeV halos as pulsar wind nebulae, despite the fact that the dynamics assumed by both groups are similar. }

\section{Associating Very High-Energy HAWC Sources With Known Pulsars}

We begin by describing the known pulsars that could potentially be responsible for powering the nine sources found in the eHWC ($>$\,$56 \, {\rm TeV}$) catalog~\cite{Abeysekara:2019gov}. In Table~\ref{TableAsso}, we list some of the selected characteristics of these pulsars, as reported in the Australia Telescope National Facility (ATNF) pulsar catalog~\cite{Manchester:2004bp}. This list of pulsars was identified in Ref.~\cite{Abeysekara:2019gov} based on their locations (within $0.5^{\circ}$ of the corresponding HAWC sources), and their high spindown power. In some cases, other nearby pulsars are not listed, primarily when observations indicate that they have substantially lower values of $\dot{E}$.

Comparing the values of the spindown luminosity of these pulsars, $\dot{E}/4\pi d^2$, to their integrated gamma-ray flux, $F_{\gamma}$, it is clear that their rotational kinetic energy is more than sufficient to produce the very high-energy gamma-ray emission reported by HAWC.\footnote{Note that the values of $F_{\gamma}$ given in Table~\ref{TableAsso} are based on an extrapolation to energies lower than those measured by HAWC, and thus may somewhat overestimate the total gamma-ray flux above 0.1 TeV.} More quantitatively, this comparison suggests that between 0.5\% and 20\% of these pulsars' spindown power goes into the production of gamma rays above 0.1 TeV (consistent with the range of values required to explain the TeV halos of Geminga and Monogem~\cite{Hooper:2017gtd,Linden:2017vvb,Sudoh:2019lav}). The only exception to this is eHWC J0534+220, which would be far brighter if the spindown power of its pulsar was transferred into gamma rays with similar efficiency. Given that this source is associated with the Crab Nebula, we do not find this result particularly surprising. In particular, the magnetic field of the Crab pulsar wind nebula is significantly stronger than that found among typical pulsar wind nebulae (or TeV halos), causing a large fraction of its spindown power to be transferred into the production of synchrotron emission~\cite{Lyutikov:2018dxm,Amato:2003kw,Meyer:2010tta,Abdalla:2019sbx,Khangulyan:2019csr}.

\begin{table*}
\resizebox{\columnwidth}{!}{
\begin{tabular}{|c|c|c|c|c|c|c|c|c|c|c|}
\hline
HAWC Source & Pulsar Candidate   & Distance & $\dot{E}$ & $\dot{E}/4\pi d^2$ & $F_{\gamma}$ & $F_{\gamma}$ / ($\dot{E}/4\pi d^2$)  & $P$ & $\dot{P}$ & $t_{c} \equiv P/2\dot{P}$ & Radio  \tabularnewline
 &    & $({\rm kpc})$ & $({\rm erg/s})$ & $({\rm TeV/cm^2/s})$  & $({\rm TeV/cm^2/s})$  &   &  $({\rm s})$ & $\times 10^{14}$ & $({\rm kyr})$ &  \tabularnewline
\hline
\hline
eHWC J0534+220  & PSR J0534+2200 & 2.00  & $4.5 \times 10^{38}$ & $5.9 \times 10^{-7}$ & $1.8 \times 10^{-10}$ & 0.0003  &0.033 & 42.1 & 1.26 & ${\rm Yes}$ \tabularnewline
\hline
eHWC J1809-193  & PSR J1809-1917 & 3.27  & $1.8 \times 10^{36}$ & $8.8 \times 10^{-10}$ & $8.5 \times 10^{-11}$ & 0.1  &0.083 & 2.55 & 51.4 & ${\rm Yes}$ \tabularnewline
\hline
--  & PSR J1811-1925 & 5.00  & $6.4 \times 10^{36}$ & $1.3 \times 10^{-9}$ & -- & 0.07  &0.065 & 4.40 & 23.3 & ${\rm No}$ \tabularnewline
\hline
eHWC J1825-134 & PSR J1826-1334  &  3.61  & $2.8 \times 10^{36}$ & $1.1 \times 10^{-9}$ & $2.3 \times 10^{-10}$ & 0.2  &0.101 & 7.53 & 21.4 & ${\rm Yes}$ \tabularnewline
\hline
--  & PSR J1826-1256  & 1.55  & $3.6 \times 10^{36}$ & $7.8 \times 10^{-9}$& -- & 0.03  & $0.110$  & 12.1 & 14.4 & ${\rm No}$ \tabularnewline
\hline
eHWC J1839-057  & PSR J1838-0537 & 2.0 & $6.0 \times 10^{36}$ & $7.8 \times 10^{-9}$& $4.1 \times 10^{-10}$ & 0.05  &0.146 & 47.2 & 4.89 & ${\rm No}$ \tabularnewline
\hline
eHWC J1842-035  & PSR J1844-0346 & 2.4  & $4.2 \times 10^{36}$ & $3.8 \times 10^{-9}$& $7.6 \times 10^{-11}$ & 0.02  & 0.113 & 15.5& 11.6 & ${\rm No}$ \tabularnewline
\hline
eHWC J1850+001  & PSR J1849-0001 & 7.00  & $9.8 \times 10^{36}$ & $1.0 \times 10^{-9}$ & $4.5 \times 10^{-11}$ & 0.05  & 0.039 & 1.42 & 43.1 & ${\rm No}$ \tabularnewline
\hline
eHWC J1907+063  & PSR J1907+0602  &  2.37  & $2.8 \times 10^{36}$ & $2.6 \times 10^{-9}$& $4.6 \times 10^{-11}$ & 0.02  &0.107 & 8.68 & 19.5 & ${\rm Yes}$ \tabularnewline
\hline
eHWC J2019+368  & PSR J2021+3651  &  1.80  & $3.4 \times 10^{36}$ & $5.5 \times 10^{-9}$& $2.7 \times 10^{-11}$ & 0.005  &0.104 & 9.57 & 17.2 & ${\rm Yes}$ \tabularnewline
\hline
eHWC J2030+412  & PSR J2032+4127  &  1.33  & $1.5 \times 10^{35}$ & $4.4 \times 10^{-10}$& $5.1 \times 10^{-11}$ & 0.1  &0.143 & 1.13 & 201 & ${\rm Yes}$ \tabularnewline
\hline
\end{tabular}
}
\caption{Properties of the pulsars potentially associated with the highest energy HAWC sources~\cite{Abeysekara:2019gov}, as reported in the ATNF Catalog~\cite{Manchester:2004bp}. Note that eHWC J1809-193 and eHWC J1825-134 each have two possible pulsar associations. When possible, we show the distance determinations as provided in the ATNF catalog, and for the cases of PSR J1838-0537, PSR J1844-0346, and PSR J1849-0001, we show those from Refs.~\cite{Pletsch:2012ct},~\cite{Wu:2017mnz}, and~\cite{Gotthelf:2010qw}, respectively. The quantity $F_{\gamma}$ is the integrated gamma-ray flux between 0.1 and 100 TeV, adopting the best-fit power law parameters as reported in Ref.~\cite{Abeysekara:2017hyn}. In the rightmost column, we report ``Yes'' if the ATNF catalog reports a detection of emission at any of 0.4, 1.2, or 2 GHz.}
\label{TableAsso}
\end{table*}

\begin{figure}
\includegraphics[width=\columnwidth,angle=0]{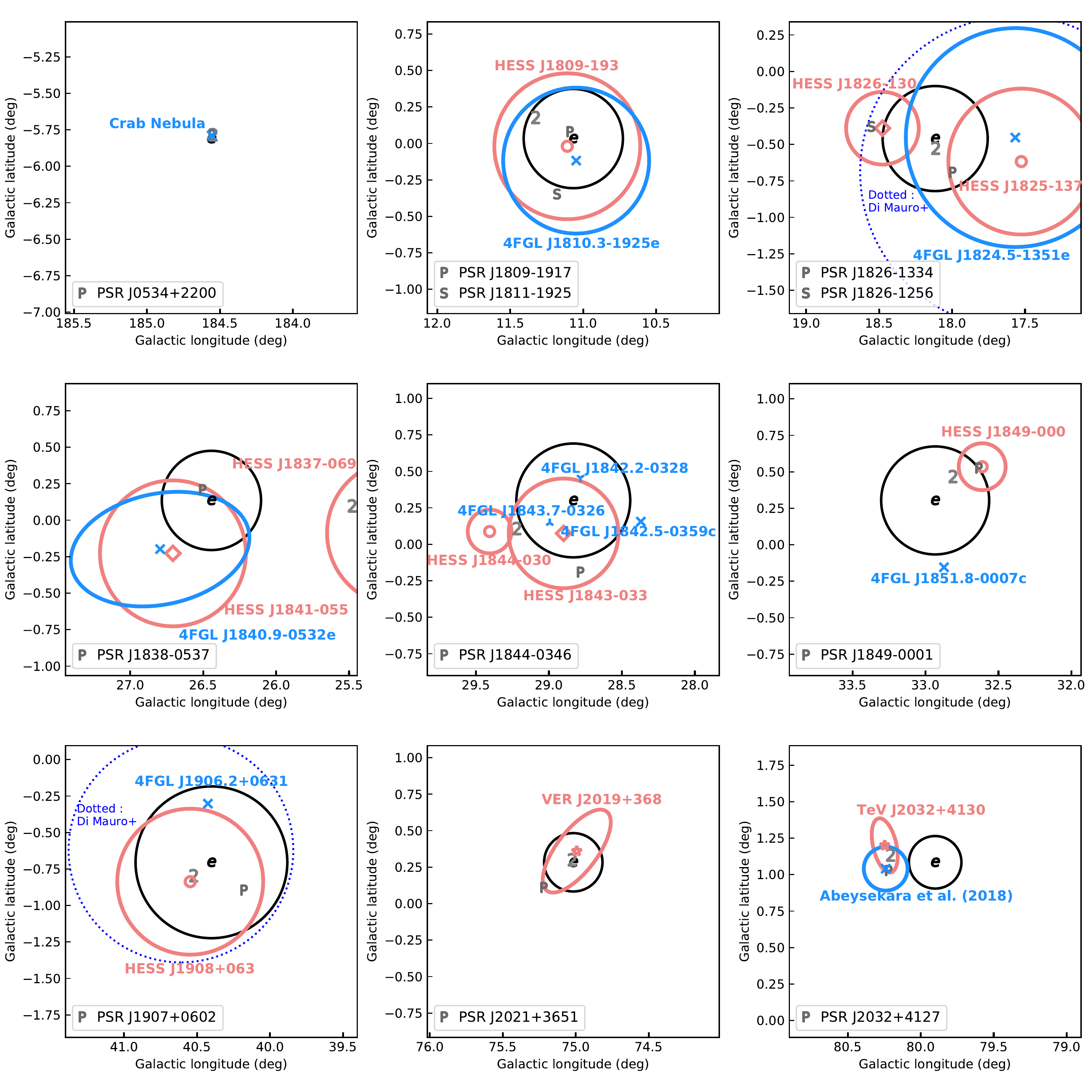}
\caption{The locations and spatial extent of the nine very high-energy HAWC sources described in Ref.~\cite{Abeysekara:2019gov}. The black `$e$' and the surrounding black circle in each frame denotes the best-fit location and 68\% containment of the source, as reported in the eHWC catalog (no circle is shown in the case of eHWC J0534+220, as its morphology is consistent with that of a point source). The symbol `2' represents the best-fit center of the source as reported in the previous 2HWC catalog. The symbol `P' (and in cases with multiple possible associations, the symbol `S') represents the location of the associated pulsars (see Table~
\ref{TableAsso}). Also shown are the location and spatial extent of any nearby TeV gamma-ray sources (red), as reported by HESS, VERITAS, and/or MAGIC, as well as the GeV counterparts as detected by Fermi (blue)~\cite{Fermi-LAT:2019yla,2018ApJ...861..134A}. The dotted blue circles represent the best-fit spatial extent of the GeV emission, as reported in Ref.~\cite{DiMauro:2020jbz}.}
\label{figmap}
\end{figure}

\begin{figure}
\includegraphics[width=\columnwidth,angle=0]{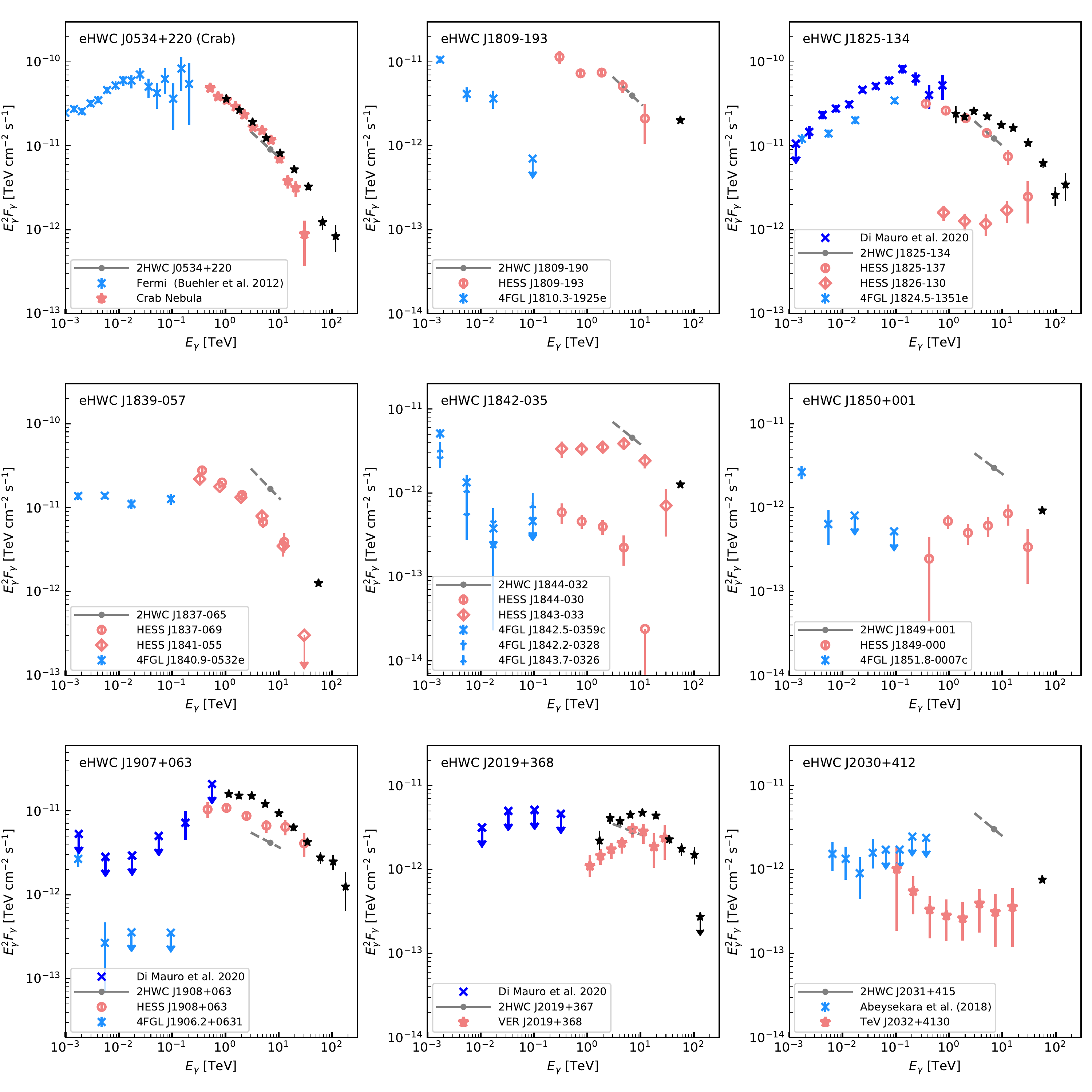}
\caption{The gamma-ray spectra of the nine very-high energy HAWC sources reported in the eHWC catalog~\cite{Abeysekara:2019gov} (black stars), as well as the best-fit power law from the earlier 2HWC catalog~\cite{Abeysekara:2017hyn} (gray dashed). Also shown are the spectra of the potential counterparts measured by HESS, VERITAS, and Fermi~\cite{Acciari:2018wql,Aliu:2014xra,Fermi-LAT:2019yla,DiMauro:2020jbz,2018ApJ...861..134A}. We note that the gamma-ray fluxes reported by different telescopes (such as HAWC and HESS) can in some cases be different, likely resulting from the larger (typically $\sim 2^{\circ}$) angular extension adopted in the analysis of the HAWC Collaboration. Note that for eHWC sources which only have an integrated flux above 56 TeV reported, we have adopted a spectral index of 2.5 in this figure.}
\label{figspec}
\end{figure}

In Fig.~\ref{figmap}, we show the locations of the nine very-high-energy HAWC sources, along with the positions of any pulsars and gamma-ray sources that are potentially associated with them. In each frame, we show the location of the HAWC source as reported in Ref.~\cite{Abeysekara:2019gov}, as well as in the earlier 2HWC catalog~\cite{Abeysekara:2017hyn}. Following Ref.~\cite{Abeysekara:2017hyn}, we also show any nearby TeV gamma-ray sources, as reported by HESS, VERITAS, and/or MAGIC. In addition, we show any GeV counterparts\footnote{We disregard Fermi-LAT sources that are identified as GeV pulsars, as the pulsed spectrum from these sources falls off rapidly above 10~GeV, and does not appreciably contribute to the TeV emission.} that are associated with a TeV source according to Fermi's 4FGL catalog~\cite{Fermi-LAT:2019yla} (unless otherwise noted below), as well as those very recently reported in Ref.~\cite{DiMauro:2020jbz}. When there is no counterpart listed in the 4FGL catalog, we include any other 4FGL gamma-ray sources that lie within 0.5$^{\circ}$ of a given eHWC source. Note that the circles shown in this figure do not represent the uncertainties pertaining to a given source's location, but rather the best-fit angular extent as reported by each collaboration. Also note that in the case of J0534+220 (the Crab Nebula), the gamma-ray emission is observed to be point-like and coincident with the radio pulsar PSR J0534+2200. While it is very likely that most of these pulsars are authentically associated with the eHWC source in question, it would not be surprising if one or two were the result of chance alignment, and not physically related to the corresponding gamma-ray source.

We will now comment on these associations on a source-by-source basis:
\begin{itemize}
    \item{eHWC J0534+220 is associated with the Crab Nebula~\cite{Abeysekara:2017hyn}. This source has been detected over a wide range of wavelengths~\cite{Lyutikov:2018dxm,Amato:2003kw,Meyer:2010tta,Abdalla:2019sbx}, and is a point-like and well-localized gamma-ray source, coincident with the radio pulsar PSR J0534+2200.}
    \item{eHWC J1809-193 is associated with the unidentified TeV gamma-ray source HESS J1809-193~\cite{Abeysekara:2017hyn}, and with the extended GeV source 4FGL J1810.3-1925e~\cite{Fermi-LAT:2019yla}.}
    \item{eHWC J1825-134 is associated with HESS J1825-137~\cite{Abeysekara:2017hyn}, which is known to be a spatially extended pulsar wind nebula. This source is also close to another source (HESS J1826-130). However, this second HESS source is dimmer than the 2HWC source by nearly an order of magnitude (see Fig.~\ref{figspec}), suggesting that the association is unlikely. This HAWC source is also associated with the extended Fermi source 4FGL J1824.5-1351e~\cite{Fermi-LAT:2019yla}. We additionally show the extent of this source as reported in Ref.~\cite{DiMauro:2020jbz}, using their ``IEM-4FGL'' interstellar emission model and the 30-100 GeV energy bin.}
    \item{eHWC J1839-057 is associated with HESS J1841-055~\cite{Abeysekara:2017hyn}, which is a complex region containing two supernova remnants, three bright pulsars, and one X-ray binary. While HESS J1837-069 is also considered as a potential counterpart to the 2HWC source~\cite{Abeysekara:2017hyn}, this HESS source is located relatively far ($\sim$\,$1^\circ$) from the best-fit position of eHWC source. Since this separation is larger than the HAWC angular resolution and extension of the eHWC source ($\sim$\,$0.3^\circ$), this association seems unlikely. HESS J1841-055 is also associated with 4FGL J1840.9-0532e~\cite{Fermi-LAT:2019yla}.}
    \item{eHWC J1842-035 is associated with the unidentified source HESS J1843-033~\cite{Abeysekara:2017hyn}. Although the 2HWC paper also considers HESS J1844-030 as a potential association~\cite{Abeysekara:2017hyn}, the flux of this HESS source is dimmer than HAWC's measurement by nearly an order of magnitude (see Fig.~\ref{figspec}), making this association unlikely. eHWC J1842-035 is located within $0.5^{\circ}$ of three Fermi sources: 4FGL J1842.5-0359c, 4FGL J1842.2-0328, and 4FGL J1843.7-0326.}
    \item{eHWC J1850+001 is associated with the pulsar wind nebula HESS J1849-000~\cite{Abeysekara:2017hyn} and is located within $0.5^{\circ}$ of the Fermi source 4FGL J1851.8-0007c.}
    \item{eHWC J1907+063 is associated with the pulsar wind nebula MGRO J1908+06~\cite{Abeysekara:2017hyn}, which is also known as HESS J1908+063, and is also associated with 4FGL J1906.2+0631~\cite{Fermi-LAT:2019yla}. We additionally show the extent of this source as reported in Ref.~\cite{DiMauro:2020jbz}, using their ``IEM-4FGL'' interstellar emission model. The flux reported in Ref.~\cite{DiMauro:2020jbz} for this source is very different from that listed in the 4FGL catalog, due to this source's significant spatial extension. In presenting our results, we will use the spectrum for this source as reported in Ref.~\cite{DiMauro:2020jbz}.}
    \item{eHWC J2019+368 is associated with the source \mbox{VER J2019+368~\cite{Abeysekara:2017hyn}}, an extended source that covers two pulsars and one star-forming region. We present the spectra of this source as reported in Ref.~\cite{Aliu:2014xra}. There are no sources in the 4FGL catalog located near eHWC 2019+368, nor is there any extended emission reported in Ref.~\cite{DiMauro:2020jbz}.}
    \item{eHWC J2030+412 is associated with the pulsar wind nebula TeV J2031+4130~\cite{Abeysekara:2017hyn}. We present the spectra reported by the MAGIC Collaboration in Ref.~\cite{Acciari:2018wql}. The flux measured by HAWC comes from a larger angular region and is much brighter than that measured by VERITAS and MAGIC, suggesting a contribution from additional components. Although there are no sources near eHWC 2030+412 in the 4FGL catalog, a potential GeV counterpart (identified using Fermi data) has been reported in Ref.~\cite{2018ApJ...861..134A}.}
\end{itemize}

In Fig.~\ref{figspec}, we show the spectra of these nine very-high-energy HAWC sources as reported in the eHWC catalog~\cite{Abeysekara:2019gov}, as well as the best-fit power law from the earlier 2HWC catalog~\cite{Abeysekara:2017hyn}. Also shown in these frames are the spectra of the potential counterparts measured by HESS, VERITAS, and Fermi. In most cases, these measurements lead to a consistent picture across a wide range of energies. We note that measurements by different telescopes (such as HAWC and HESS) can in some cases be different, due to the treatment of these sources' spatial extension.

\section{Pulsars and Inverse Compton Emission}
\label{sec:calc}

In this section, we describe our calculation of the gamma-ray spectrum produced through the inverse Compton scattering of a population of very high-energy electrons and positrons, injected with a given spectrum and over a given time profile.

Very high-energy electrons and positrons lose energy through a combination of inverse Compton scattering and synchrotron processes, leading to the following energy loss rate~\cite{Blumenthal:1970gc}:
\begin{eqnarray}
-\frac{dE_e}{dt} &=& \sum_i \frac{4}{3}\sigma_T u_i S_i(E_e) \bigg(\frac{E_e}{m_e}\bigg)^2 + \frac{4}{3}\sigma_T u_{\rm mag} \bigg(\frac{E_e}{m_e}\bigg)^2  \nonumber \\
&\equiv& b(E_e) \,  \bigg(\frac{E_e}{{\rm TeV}}\bigg)^2,
\end{eqnarray}
where $\sigma_T$ is the Thomson cross section and
\begin{eqnarray}
b &\approx&  1.02 \times 10^{-13} \, {\rm TeV}/{\rm s} \, \nonumber \\
&\times& \bigg[ \sum_i \frac{u_{i}}{{\rm eV}/{\rm cm}^3} \, S_{i}(E_e) + \frac{u_{\rm mag}}{{\rm eV}/{\rm cm}^3} \bigg].
\end{eqnarray}
The sum in this expression is carried out over the various components of the radiation backgrounds, consisting of the cosmic microwave background (CMB), infrared emission (IR), and starlight (star). We take each of these radiation components to have a blackbody spectrum and adopt the following values for their energy densities and temperatures: $u_{\rm CMB}=0.260$ eV/cm$^3$, $u_{\rm IR}=0.30$ eV/cm$^3$, $u_{\rm star}=0.3$ eV/cm$^3$, $T_{\rm CMB} =2.7$ K, $T_{\rm IR} =20$ K, and $T_{\rm star} =5000$ K~\cite{Porter:2017vaa}. For the energy density of the magnetic field, we adopt as our default value $u_{\rm mag}=0.224$ eV/cm$^3$, corresponding to $B\simeq 3\,\mu$G. At relatively low electron energies, these parameters correspond to a value of $b \simeq 1.2 \times 10^{-13}$ TeV/s ($S_i \approx 1$). At very high energies ($E_e \gsim m^2_e/2T$), however, the inverse Compton scattering will be well outside of the Thomson regime, and Klein-Nishina suppression will play an important role. For our calculations, we utilize the full Klein-Nishina cross-section formula, as calculated in Ref.~\cite{Blumenthal:1970gc} and as implemented in the publicly available code {\sc naima}~\cite{naima} (see also Refs.~\cite{2010PhRvD..82d3002A,2014ApJ...783..100K}). For illustrative purposes, however, the key features of Klein-Nishina suppression can be seen more clearly using the following approximate expression~\cite{schlickeiser2010}:
\begin{equation}
S_i (E_e) \approx \frac{45 \, m^2_e/64 \pi^2 T^2_i}{(45 \, m^2_e/64 \pi^2 T^2_i)+(E^2_e/m^2_e)}.
\end{equation}

For electrons of a given energy, the effects of Klein-Nishina suppression are most pronounced for the highest-energy target photons. For the very-high-energy ($E_e\gsim\,{\rm TeV}$) electrons/positrons of most interest to this study, energy losses from inverse Compton scattering are dominated by scattering with the IR background, as well as the CMB. At energies greater than $\sim$\,$50 \, {\rm TeV}$, the CMB alone dominates this process.

Over a period of time in which an electron or positron of energy $E_e$ loses a small quantity of energy, $\Delta E_e$, that particle will generate the following spectrum of inverse Compton emission:
\begin{eqnarray}
\frac{dN_{\gamma}}{dE_{\gamma}}(E_{\gamma}, E_e) &=& A(E_e,\Delta E_e) \, f_{\rm ICS}(E_e) \, l_{e} \\
&\times& \int    \frac{dn}{d\epsilon}(\epsilon) \, \frac{d\sigma_{ICS}}{dE_{\gamma}}(\epsilon,E_{\gamma},E_{e}) \,   d\epsilon,
\nonumber
\end{eqnarray}
where $dn/d\epsilon$ is the spectrum of target radiation, which we take to the be the sum of the blackbody distributions described above. The quantity $A$ is set by requirement that $\Delta E_e = \int dE_{\gamma} \, E_{\gamma} \, dN_{\gamma}/dE_{\gamma}$, and $f_{\rm ICS}(E_e)$ is the fraction of the electron or positron's energy losses that are from inverse Compton scattering (as opposed to synchrotron). The differential cross section for inverse Compton scattering is given by~\cite{aharonian1981}:
\begin{eqnarray}
\frac{d\sigma_{ICS}}{dE_{\gamma}}(\epsilon,E_{\gamma},E_{e})&=&\frac{3\sigma_{T}m_{e}^{2}}{4\epsilon E_{e}^{2}}\,
\bigg[ 1+\bigg( \frac{z^{2}}{2(1-z)}\bigg)  \\
+\bigg(\frac{z}{\beta (1-z)}\bigg)&-&\bigg(\frac{2z^{2}}{\beta^{2}(1-z)}\bigg)
-\bigg(\frac{z^{3}}{2\beta (1-z)^{2}}\bigg) \nonumber \\
&-&\bigg(\frac{2z}{\beta (1-z)}\bigg)\ln\bigg(\frac{\beta (1-z)}{z}\bigg) \bigg], \nonumber
\end{eqnarray}
where $z \equiv E_{\gamma}/E_{e}$ and $\beta \equiv 4\epsilon E_{e}/m_{e}^{2}$. At energies within the range measured by HAWC, inverse Compton scattering generally yields photons with energies not very far below that of the incident electrons and positrons, $E_{\gamma} \sim E_e$.

As time passes, pulsars slow down and lose rotational kinetic energy, transferring much of this energy into the acceleration of particles which produce the radio, gamma-ray, and other non-thermal emission that is observed from these objects. From the measured quantities $P$ and $\dot{P}$, we can define the pulsar's characteristic age, $t_c$:
\begin{equation}
    t_{c} \equiv \frac{P}{2\dot{P}} = \frac{n-1}{2}(t_{\rm age} + \tau),
\end{equation}
where $n$ is the braking index, $t_{\rm age}$ is the age of the pulsar, and $\tau$ is its spindown timescale. From the spindown equations, we can write the spindown timescale as
\begin{equation}
    \tau = \frac{2t_{c}}{n-1}\left(\frac{P_0}{P}\right)^{n-1},
\end{equation}
where $P_0$ is the initial period of the pulsar. For a given set of $P_0$ and $n$, these equations determine $\tau$ and $t_{\rm age}$. The spindown power of a pulsar evolves as follows:
\begin{equation}
\dot{E}(t) = 4\pi^2 I \frac{\dot{P}}{P^3} = \dot{E}_0\left(1 + \frac{t}{\tau}\right)^{-\frac{n+1}{n-1}},
\end{equation}
where $\dot{E}_0$ is the initial spindown power, given by
\begin{equation}
\dot{E}_0  = 4\pi^2 I \frac{\dot{P}}{P^3}\left(1 + \frac{t_{\rm age}}{\tau}\right)^{\frac{n+1}{n-1}}.
\end{equation}

These equations leave us with $P_0$, $n$, and $I$ as free parameters. Unless otherwise stated, we will adopt $I=10^{45}$~g~cm$^2$ and $n=3$ throughout this study.

\section{Results}

\begin{figure}
\includegraphics[width=\columnwidth,angle=0]{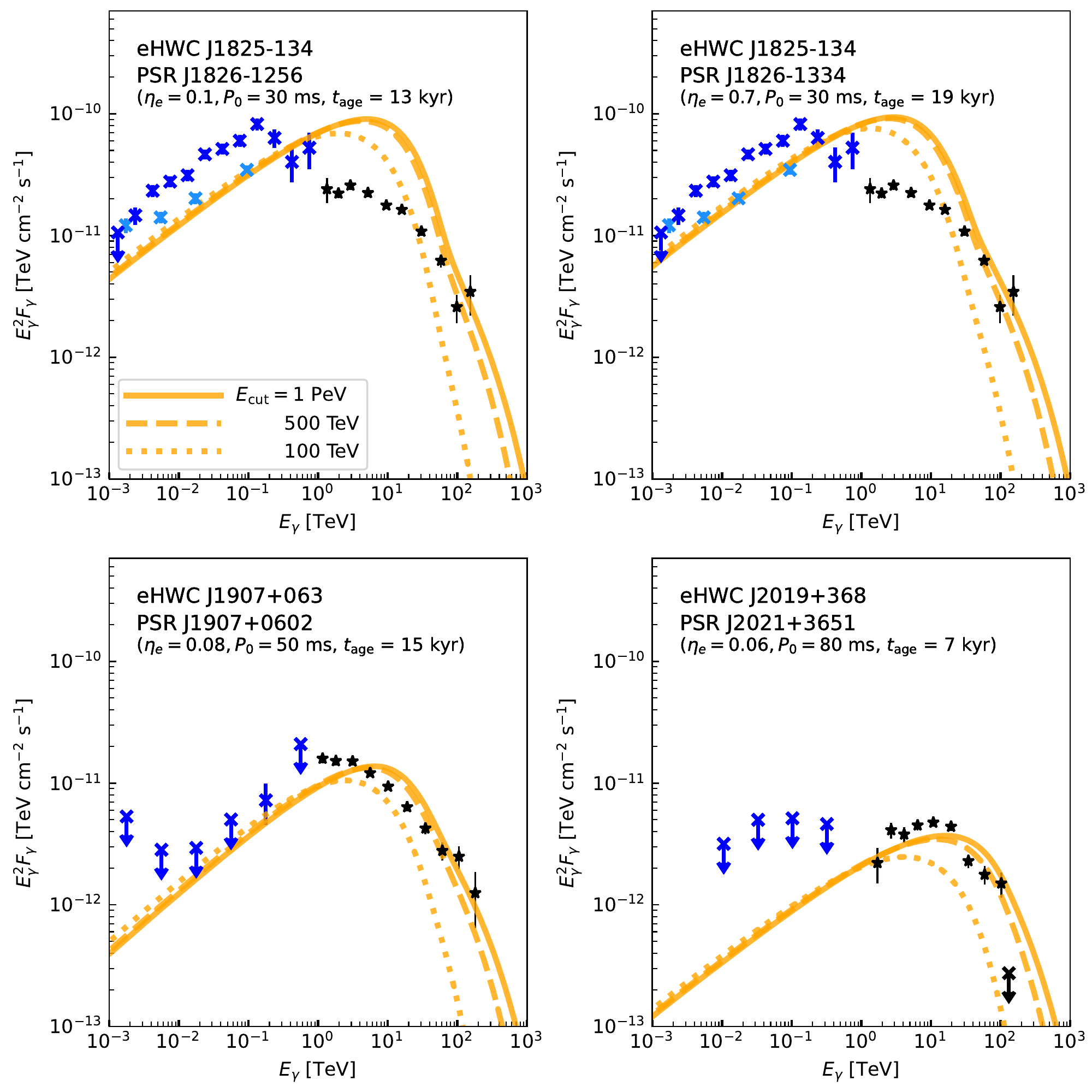}
\caption{The spectrum of inverse Compton emission predicted from the pulsars PSR J1826-1256, PSR J1826-1334, PSR J1907+0602, and PSR J2021+3651, compared to the measured gamma-ray spectra from the associated HAWC sources (see Fig.~\ref{figspec}). We have parameterized the injected electron/positron spectrum as $dN_e/dE_e \propto E_e^{-\gamma} \exp(-E_e/E_{\rm cut})$ and show results for three values of $E_{\rm cut}$. We adopt a spectral index of $\gamma=2.1$ in this figure, except for in the bottom left frame, where we have used $\gamma=2.0$. For each pulsar, we have selected values for the electron/positron efficiency ($\eta_e$) and initial period ($P_0$) which lead to reasonable agreement with the observed spectrum and intensity of each source. We also show the value of each pulsar's age, as calculated from the value of $P_0$. We emphasize that the cutoff observed in these spectra above $\sim$\,$1-10 \, {\rm TeV}$ is an unavoidable consequence of the model, and occurs when the age of a pulsar exceeds the timescale for electron/positron energy losses.}
\label{fig1}
\end{figure}

In Fig.~\ref{fig1}, we show the spectra of inverse Compton emission predicted from the pulsars PSR J1826-1256, PSR J1826-1334, PSR J1907+0602, and PSR J2021+3651, comparing our results with the gamma-ray observations of each associated HAWC source. In each case, we have parameterized the injected electron/positron spectrum as a power-law with an exponential cutoff, $dN_e/dE_e \propto E_e^{-\gamma} \, \exp(-E_e/E_{\rm cut})$. Along with $\gamma$ and $E_{\rm cut}$, we treat as free parameters each pulsar's initial period, and the fraction of its spindown power that goes into the production of electrons and positrons integrated above 10 GeV, $\eta_e$. For each pulsar's distance, period, and rate of change of its period, we adopt the values reported in the Australia Telescope National Facility (ATNF) pulsar catalog~\cite{Manchester:2004bp} (as shown in Table~\ref{TableAsso}). We adopt $\gamma=2.0$ in the bottom left frame of Fig.~\ref{fig1}, and 2.1 in the other three frames. In each frame, we show results for three choices of $E_{\rm cut}$, and have selected values of $\eta_e$ and $P_0$ (obtaining the corresponding value of $t_{\rm age}$) which, when possible, lead to reasonable agreement with the observed spectral shape and intensity of each source.

\begin{figure}
\includegraphics[width=\columnwidth,angle=0]{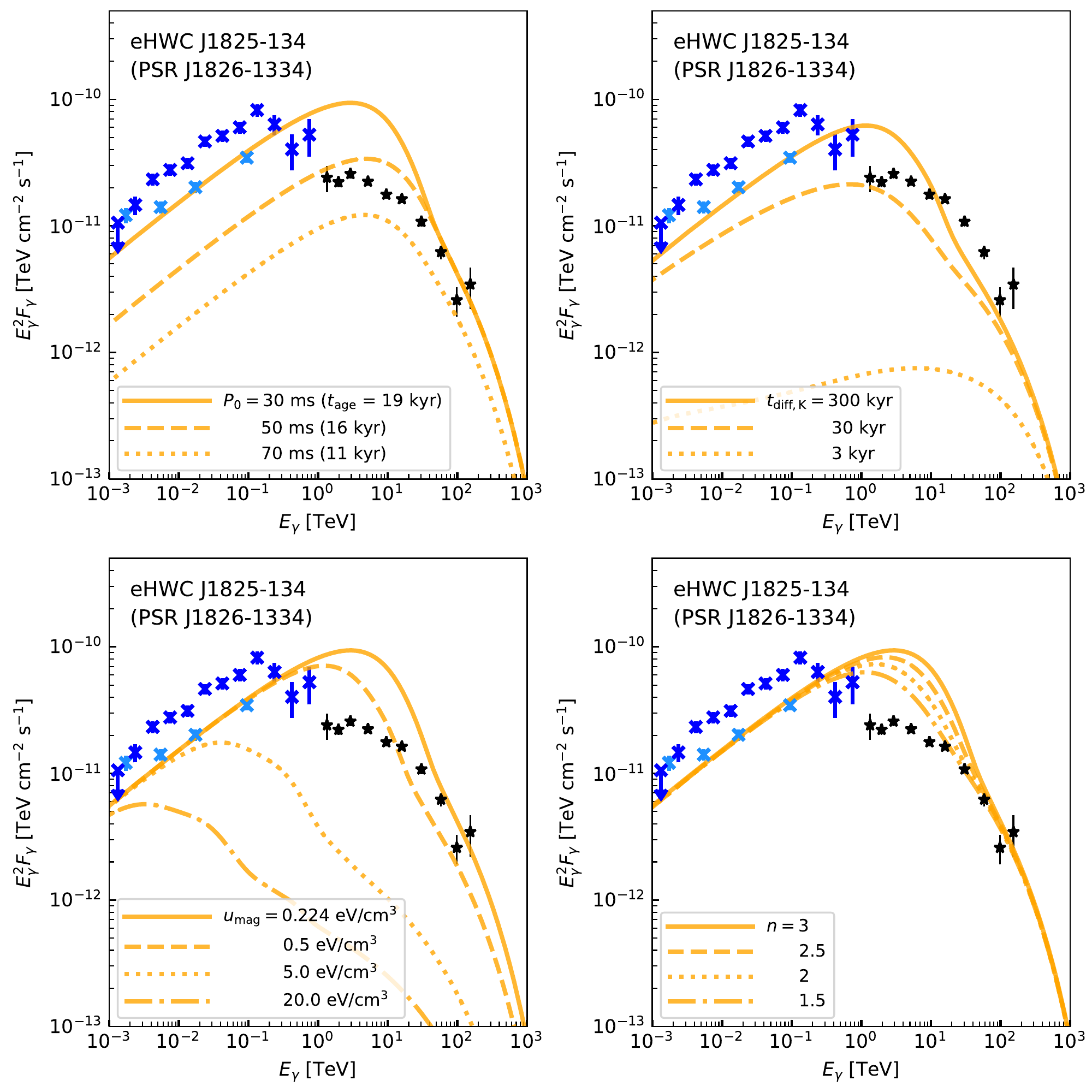}
\caption{As in Fig.~\ref{fig1}, but for selected parameter variations, and focusing on the case of eHWC J1825-134/PSR J1826-1334. In the upper left frame, we consider three different combinations for the values of the initial period and pulsar age, while in the lower left frame we show results for three choices of the energy density of the magnetic field. In the upper right frame, we consider scenarios in which the electrons and positrons are able to escape the emitting region on a timescale given by \mbox{$t_{\rm esc}=t_{\rm diff, K} \times (E_{e}/{\rm TeV})^{-1/3}$}, corresponding to the energy dependence predicted for Kolmogorov diffusion. In the lower right frame, we show results for four choices of the pulsar's braking index, $n$. In each panel, unless stated otherwise, we have adopted $P_0=30 \, {\rm ms}$ ($t_{\rm age}=19 \, {\rm kyr}$), and $\eta_{e}=0.7$.}
\label{fig2}
\end{figure}

As seen in Fig.~\ref{fig1}, the gamma-ray spectrum that is produced through inverse Compton scattering is automatically suppressed at energies above $\sim$\,$10 \,{\rm TeV}$, for which the age of these pulsars exceeds the timescale for electron/positron energy losses, $t_{\rm age} \gsim (bE_e)^{-1}$. Around this energy, the spectrum of the ambient electrons and positrons transitions from the injected index ($\gamma$) to a significantly softened index ($\gamma-1$). Note that this suppression occurs even if the injected spectrum does not have a cutoff in the relevant energy range ($E_{\rm cut} \gg 100 \, {\rm TeV}$). Klein-Nishina effects also influence the exact shape of the high-energy spectrum. In these results, there is no indication of a cutoff in the injected spectrum of electrons and positrons, suggesting that these sources accelerate such particles to at least several hundred TeV. At lower energies, $E_e \ll (b \, t_{\rm age})^{-1} \sim 20-50 \, {\rm TeV}$, the electrons and positrons that have been injected from the pulsar over its history have not lost much of their initial energy. In this limit, the normalization of the gamma-ray spectrum is set by the total integrated energy that has been injected from the pulsar in the form of electrons and positrons, which is proportional to $\eta_e \, t_{\rm age}/P^2_{0}$.

From the lower frames of Fig.~\ref{fig1}, we see that the pulsars PSR J1907+0602 and PSR J2021+3651 can produce the emission observed by HAWC and Fermi, in each case requiring an efficiency similar to that of Geminga or Monogem, $\eta_e \sim 0.1$. In the upper frames, we see that either PSR J1826-1256 or PSR J1826-1334 (or some combination thereof) could be responsible for the gamma-ray emission attributed to eHWC J1825-134, although the latter would require a high value of $\eta_e \sim 0.7$, and neither of these pulsars provides a particularly good fit in the $\sim 1-10$ TeV range.

In Fig.~\ref{fig2}, we consider some variations regarding our parameter choices, focusing on the case of eHWC J1825-134 and its corresponding PSR J1826-1334. In the upper left frame, we consider three different combinations for the values of the initial period and pulsar age. As described above, this does not impact the spectrum at high energies, where only the current power of the injected electrons/positrons determines the normalization. At lower energies, however, the normalization scales as $\eta_e t_{\rm age}/P^2_{0}$, corresponding to the total energy injected into high-energy electrons and positrons over the life of the pulsar. In the lower-left frame, we consider variations to the energy density of the magnetic field, showing results for $u_{\rm mag}=0.224$~eV/cm$^{3}$ (our default value), 0.5~eV/cm$^{3}$, 5.0~eV/cm$^{3}$ and 20~eV/cm$^{3}$, corresponding to $B=3.0$~$\mu$G, 4.5~$\mu$G, 14.2 $\mu$G and 28.3~$\mu$G, respectively. By increasing the energy density of the magnetic field, a larger fraction of the energy in electrons and positrons is lost to synchrotron, suppressing the gamma-ray emission that is produced through inverse Compton scattering.

Thus far in our calculations, we have assumed that the electrons and positrons remain within the TeV halo or pulsar wind nebula, and do not escape via diffusion. This corresponds to one or both of the following conditions being satisfied: $t_{\rm esc} \gg t_{\rm age}$ or $t_{\rm esc} \gg (b E_e)^{-1}$, where $t_{\rm diff}$ is the timescale for particles to escape the TeV halo via diffusion. In the upper right frame of Fig.~\ref{fig2}, we consider a class of scenarios in which the electrons/positrons instead escape on a timescale given by $t_{\rm esc}=t_{\rm diff, K} \times (E_{e}/{\rm TeV})^{-1/3}$, corresponding to the energy dependence predicted for Kolmogorov diffusion. More quantitatively, we reduce the number of electrons and positrons within the emission region by a factor of $e^{-\delta t/t_{\rm esc}}$ in each timestep of length $\delta t$. The impact of diffusion is significant only when $t_{\rm esc}$ is smaller than both the age of the pulsar (which, in this case, is 19 kyr), and the timescale for energy losses (which is $\sim$\,$10^3 \, {\rm yr}$ at the highest energies shown, and $\sim$\,$10^5 \, {\rm yr}$ at TeV-scale energies). This could, in principle, significantly suppress the predicted gamma-ray emission, but only in scenarios with very rapid diffusion (much faster than favored by the spectra of Geminga and Monogem~\cite{Hooper:2017gtd}). We do not expect diffusion to play an important role in most of the sources under consideration in this study.

Lastly, in the lower right frame of Fig.~\ref{fig2}, we show results for four choices of the pulsar's braking index, $n$. The spectrum of this particular source is somewhat better fit for lower values of the braking index.

\begin{figure}
\includegraphics[width=\columnwidth,angle=0]{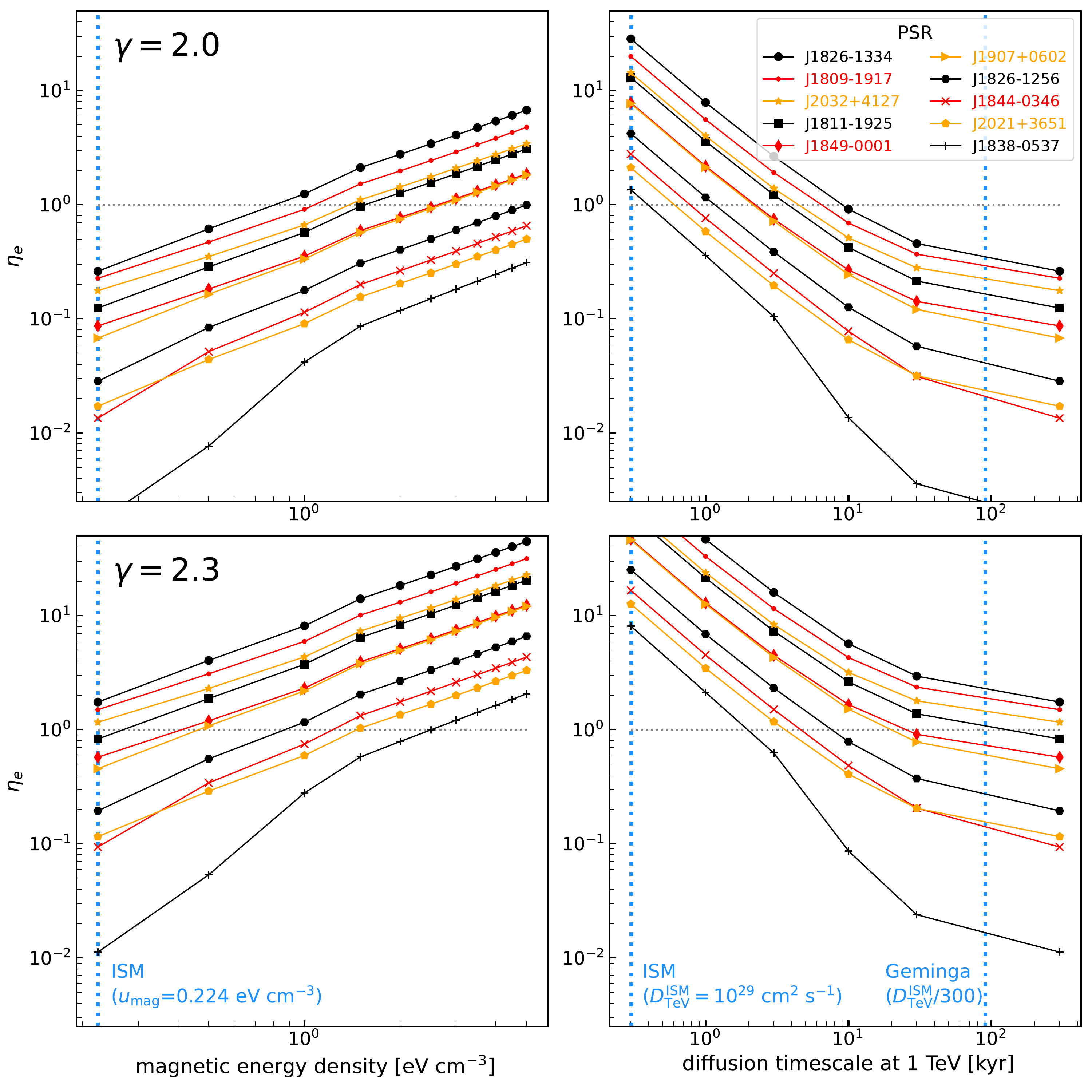}
\caption{The fraction of each pulsar's spindown power that must go into the production of electrons and positrons, $\eta_e$, in order to explain the intensity of the gamma-ray emission observed by HAWC for each of the pulsars potentially associated with a source in the eHWC catalog. These results are shown for two choices of the injected spectral index, and as a function of either the energy density of the magnetic field, or the timescale for diffusion. In each case, we have adopted $P_0=30$ ms, and $E_{\rm cut}=1$ PeV. For reference, we show as vertical lines the values of these quantities as measured in the local interstellar medium, assuming a 10~pc emitting region for our calculation of the diffusion timescale. From these results, it is clear that the intensity of the observed gamma-ray emission can be produced for reasonable efficiencies, $\eta_e \sim \mathcal{O}(0.1)$, so long as 1) the magnetic field is not much stronger than in the local interstellar medium ($u_{\rm mag} \lsim 1$ eV/cm$^{3}$), 2) diffusion is highly suppressed in the region of inverse Compton scattering ($t_{\rm diff} \gsim 10$ kyr), and 3) the injected spectral index is somewhat hard ($\gamma \sim 2$). These characteristics are consistent with those observed from the Geminga and Monogem TeV halos.}
\label{fig20}
\end{figure}

In Fig.~\ref{fig20}, we show the values of $\eta_e$ that are required to explain the intensity of the gamma-ray emission observed by HAWC from each of the sources in the eHWC catalog, for each of the potentially associated pulsars listed in Table~\ref{TableAsso} (with the exception of the Crab Pulsar, which requires a significantly smaller value of $\eta_e$ for a given value of $u_{\rm mag}$). We show results for two choices of the injected spectral index ($\gamma=2.0$, 2.3), and present these results as a function of either the energy density of the magnetic field, or the timescale for diffusion. In each case, we have adopted $P_0=30$ ms, and $E_{\rm cut}=1$ PeV. For reference, we show as vertical lines the values of these quantities as measured in the local interstellar medium.

From Fig.~\ref{fig20}, it is clear that in the case of $\gamma=2$, the intensity of the observed gamma-ray emission can be produced for reasonable efficiencies, $\eta_e \sim \mathcal{O}(0.1)$, so long as 1) the magnetic field is not much stronger than in the local interstellar medium ($u_{\rm mag} \lsim 1$ eV/cm$^{3}$), and 2) diffusion is highly suppressed in the region of inverse Compton scattering ($t_{\rm diff} \gsim 10$ kyr), as is known to be the case for both the Geminga and Monogem TeV halos. Comparing this to the results found in the $\gamma=2.3$ case, it is clear that somewhat hard spectral indices are also required to produce the observed emission, again consistent with that observed from Geminga and Monogem. Note that in calculating the values of $\eta_e$, we have adopted a power-law injected spectrum of electrons and positrons, integrated to a minimum energy of 10 GeV. Multiwavelength studies of pulsar wind nebulae often require the electrons/positrons to be injected with a broken power-law spectrum, with \mbox{$E_{\rm br} \sim$ 0.1~TeV} (see, for example, Ref.~\cite{Torres:2014iua}). Adopting such a function can reduce the required efficiency by a factor of approximately $\sim (E_{\rm br}/10~\rm GeV)^{\gamma-2}$.

\section{Comparison of Hadronic and Leptonic Models}

\begin{figure}
\includegraphics[width=\columnwidth, angle=0]{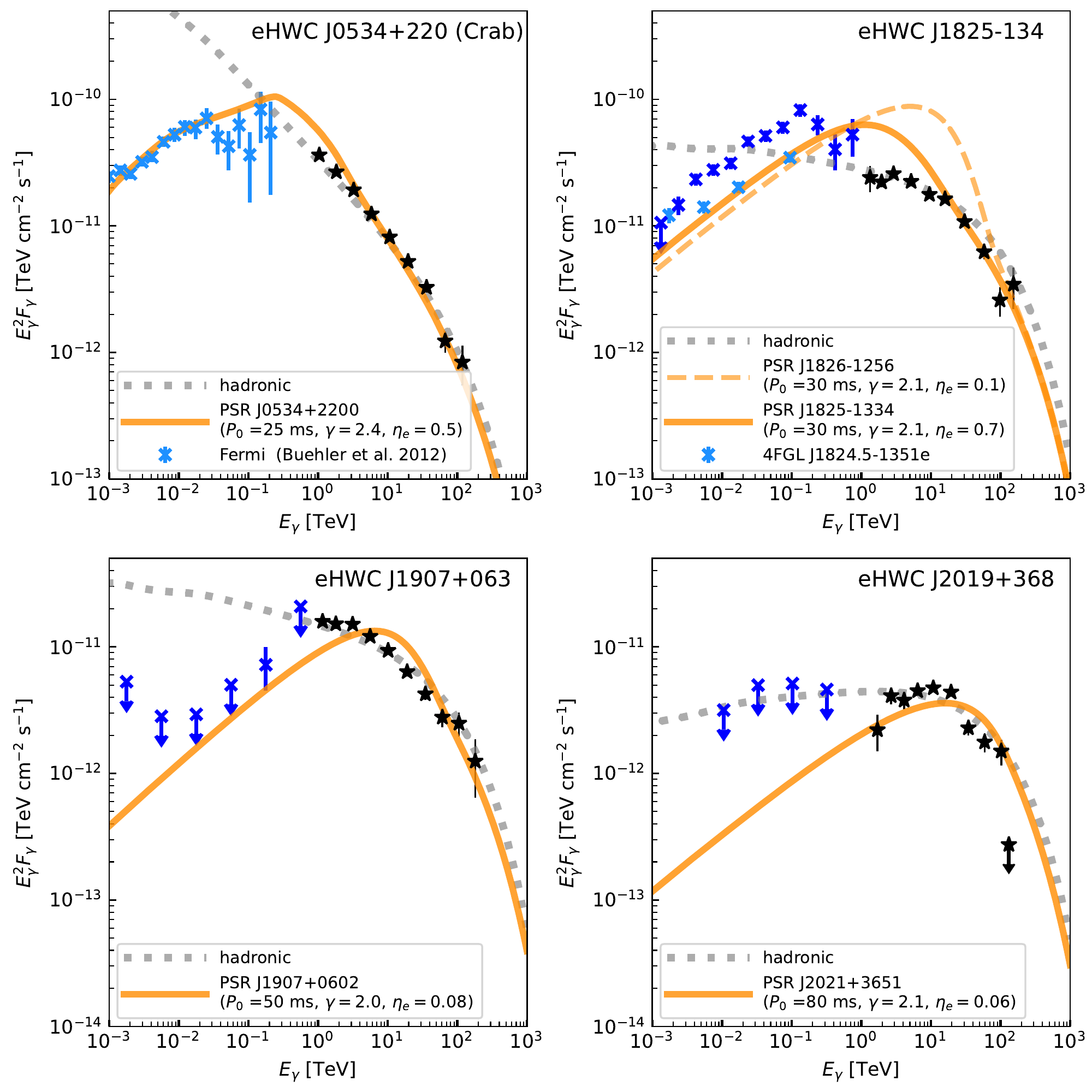}
\caption{A comparison of the gamma-ray spectra observed from four of the sources in the eHWC catalog to that predicted from both leptonic and hadronic models. For three of these four sources, the hadronic emission that is predicted at GeV-scale energies significantly exceeds that observed by Fermi. We have adopted a spectrum of protons that is described by a power-law with an exponential cutoff, \mbox{$dN/dE \propto E^{-p} \exp(-E/E_{\rm cut})$}, with $E_{\rm cut} =1$ PeV and extending to a minimum energy of 10 GeV. In each frame, we have adopted a value of $p$ which best accommodates the spectra reported by HAWC.}
\label{fig4}
\end{figure}

The spectral feature that is observed from these sources around $\sim$1-10 TeV is a natural consequence of our leptonic model, and occurs at the energy where the timescale for energy losses matches the age of the pulsar. In hadronic models, no such feature is expected. Furthermore, hadronic models that can explain the spectrum observed from these sources at very high energies generally predict far more emission than is observed in the GeV range~\cite{Fermi-LAT:2019yla,DiMauro:2020jbz,2018ApJ...861..134A}. This disfavors models in which these sources produce their gamma-ray emission primarily through hadronic processes.

In Fig.~\ref{fig4}, we compare the gamma-ray spectra observed from four of the sources in the eHWC catalog to that predicted in both leptonic and hadronic models. As we did to calculate the emission from inverse Compton scattering, we made use of the publicly available code {\sc naima} to determine the spectra of hadronic gamma-ray emission~\cite{naima} (see also Refs.~\cite{2014PhRvD..90l3014K}). For three of these four sources, the hadronic emission that is predicted at GeV-scale energies significantly exceeds that observed by Fermi. In this figure, we have adopted a spectrum of protons that is described by a power-law with an exponential cutoff, $dN/dE \propto E^{-p} \exp(-E/E_{\rm cut})$, extending from a minimum energy of 10 GeV, and with $E_{\rm cut} =1$ PeV. In each frame, we have adopted values of $p$ which best accommodate the spectrum reported by HAWC ($p=2.65$ in the upper left, 2.15 in the upper right, 2.2 in the bottom left, and 2.0 in the bottom right). In the case of the Crab Nebula (upper left), we adopt a braking index of $n=2.5$ in the leptonic model, and adopt a large value for the strength of the magnetic field, $B=90~\mu$G (in order to be compatible with the emission observed in the $\sim \mathcal{O}(0.1 \,{\rm GeV})$ range, which is attributed to synchrotron). In the case of the Crab Nebula, we have also included synchrotron photons as targets of inverse Compton scattering, within a region taken to be 2 parsecs in radius. For PSR J1825-1334 we have adopted $n=1.5$, while we have retained our default choice of $n=3$ for the three remaining pulsars. For each curve, we adopt a normalization to match the HAWC data.

Since each eHWC source is located in a region where young pulsars, and hence recent star formation and supernova explosion exist, there may also be contributions from hadronic processes to the gamma-ray flux. Comparing our model curves with GeV data indicates that, hadronic component could also produce a significant fraction of the observed flux for eHWC J2019+368, while likely at most $\sim$10$\%$ level for the other sources. We note that our hadronic models assume that protons are injected with a single power law. If we assume a broken power law to reduce the energy injected into GeV-scale protons, more contributions from hadronic processes could be allowed without violating the Fermi data. Such a hard spectrum could be realized in a scenario where very-high-energy protons that escape early from the SNR travel into massive gas clouds, producing gamma rays there, while lower-energy protons remain confined in the accelerator~(e.g., \cite{2007ApJ...665L.131G}). However, the eHWC sources shown in Figure~\ref{fig4}, except for eHWC 1825-134, have not been reported to have a clear spatial correlation with gases, which challenges this scenario. Regarding eHWC 1825-134, mixed hadronic/leptonic contributions is a plausible scenario. (\cite{Albert:2020yty}, see also Sec. \ref{sec:summary})

\section{Discussion and Summary}
\label{sec:summary}
The nature of the highest energy HAWC sources is a subject of considerable interest, which has recently been discussed by a number of authors and collaborations.  In particular, the HAWC Collaboration has used multiwavelength data to argue that the gamma-ray emission from eHWC J2019+368 is leptonic in origin~\cite{Albert:2021uyy}, in agreement with our assessment of this source. More recently, HAWC has performed a stacking analysis of ten pulsars that are not associated with any eHWC sources, identifying evidence of gamma-ray emission at energies above 56 TeV~\cite{Albert:2021vrd}. More broadly speaking, they conclude from this information that high-spindown power pulsars universally produce extremely high energy photons. In Ref.~\cite{Albert:2020yty}, members of the HAWC Collaboration argued that eHWC J1825-134 can be separated into four components: diffuse Galactic emission, HAWC J1826-128 (the counterpart to HESS J1826-130), HAWC J1825-138 (the counterpart to HESS J1825-137), and the newly discovered source HAWC J1825-134. The spectrum of the emission associated with HAWC J1825-134, and its spatial correlation with dense gas, favors a hadronic interpretation for this emission. In contrast, the other two HAWC sources that contribute to eHWC J1825-134 are likely leptonic in origin.

Beyond the HAWC Collaboration, Di Mauro~{\it et al.}~\cite{DiMauro:2020jbz} have shown that the spectra of three eHWC sources (eHWC J1825-134, J1907+063, and J2019+368) can be well fit by leptonic models, in concordance with our conclusions (see also, Ref~\cite{Breuhaus:2020mof}). On similar grounds, Fang~{\it et al.}~\cite{Fang:2020uiz} have argued that eHWC J2019+368 is likely leptonic in nature. In contrast, the authors of Ref.~\cite{Araya:2018zsf} have claimed that HESS J1809-193 (associated with eHWC J1809-193) is likely to be a hadronic source.

In this paper, we have studied each of the nine gamma-ray sources contained in the eHWC catalog, expanding on the previous work described above, and identifying significant evidence that their emission is likely leptonic in origin. In particular, the gamma-ray emission from these sources can be straightforwardly accommodated within a model in which $\sim \mathcal{O}(10\%)$ of the host pulsar's spindown power is transferred into the acceleration of electrons and positrons with a simple power-law spectrum. The spectral break that is observed among these sources is an unavoidable consequence of this model.

In contrast, the spectral feature that is observed from these sources is not expected in hadronic scenarios, which also predict far more emission at GeV-scale energies than is observed. For the three eHWC sources with detailed spectral information, we can rule out scenarios in which a significant fraction of their observed emission is hadronic in origin. While it remains possible that one or more of the other six eHWC sources could produce hadronic emission (see, for example, Ref.~\cite{Araya:2018zsf}), we stress that nothing in our analysis differentiates any of these sources from those that are clearly leptonic in nature. This disfavors an interpretation of these sources as the long-sought-after Galactic PeVatrons.

Furthermore, all nine sources in the eHWC catalog can be powered by the rotational kinetic energy of their host pulsar, requiring efficiencies that are similar to those of the Geminga and Monogem TeV halos. Also like Geminga and Monogem, diffusion appears to be suppressed within the emission regions of these sources, and electrons and positrons are injected into these regions with a relatively hard spectral index, $\gamma \sim 2$.

In light of the considerations described in the paragraphs above, we conclude that HAWC's highest energy sources are likely to be TeV halos or pulsar wind nebulae, which produce their gamma-ray emission through inverse Compton scattering, and which are powered by the rotational kinetic energy of their host pulsar. We find no evidence that this class of sources produces significant gamma-ray emission through hadronic processes, or accelerates protons to PeV-scale energies.

\bigskip
\bigskip
\bigskip

\textbf{Acknowledgments.} We would like to thank Mattia Di Mauro for providing us with the data from Ref.~\cite{DiMauro:2020jbz}. TS is supported by a Research Fellowship of Japan Society for the Promotion of Science (JSPS) and by JSPS KAKENHI Grant No.\ JP 18J20943. TL is partially supported by the Swedish Research Council under contract 2019-05135, the Swedish National Space Agency under contract 117/19 and the European Research Council under grant 742104. DH is supported by the Fermi Research Alliance, LLC under Contract No. DE-AC02-07CH11359 with the U.S. Department of Energy, Office of High Energy Physics. In this work, we have made use of {\sc naima}~\cite{naima}, {\sc gammapy}\footnote{https://www.gammapy.org}\cite{gammapy:2017, gammapy:2019}, {\sc astropy}\footnote{http://www.astropy.org}~\cite{astropy0,astropy}, {\sc matplotlib}~\cite{matplotlib}, {\sc numpy}~\cite{numpy}, and {\sc scipy}~\cite{scipy}.

\bibliographystyle{JHEP}
\bibliography{main_h}
\end{document}